\documentclass[prl,twocolumn,superscriptaddress]{revtex4}

\usepackage{acronym}
\usepackage{amsmath}
\usepackage{amssymb}
\usepackage{graphicx}
\def\TMC{T_{MC}}
\def\xd{\xi_{\rm d}}
\def\xs{\xi_{\rm s}}

\newcommand{\beq}{\begin{equation}}
\newcommand{\eeq}{\end{equation}}
\newcommand{\beqa}{\begin{eqnarray}}
\newcommand{\eeqa}{\end{eqnarray}}

\begin{document}

%%
%% Title, authors, institutions
%%

\title{A phase-separation perspective on dynamic heterogeneities in glass-forming liquids}

\author{C. Cammarota\footnote{Present address: CEA, Institut de Physique Theorique, Saclay, F-91191 Gif-sur-Yvette, France}} 

  \affiliation{Dipartimento di Fisica, Universit{\`a} di Roma
  ``Sapienza'', p.le Aldo Moro 5, 00185, Roma, Italy. }
  \affiliation{Centre for Statistical Mechanics and Complexity (SMC), CNR-INFM.}

\author{A. Cavagna}  
  \affiliation{Centre for Statistical Mechanics and Complexity (SMC), CNR-INFM.}
  \affiliation{Istituto Sistemi Complessi (ISC), CNR, Via dei Taurini 19, 00185 Roma, Italy.}

\author{I. Giardina}  
  \affiliation{Centre for Statistical Mechanics and Complexity (SMC), CNR-INFM.}
  \affiliation{Istituto Sistemi Complessi (ISC), CNR, Via dei Taurini 19, 00185 Roma, Italy.}

\author{G. Gradenigo\footnote{Present address: SMC-INFM and Dipartimento di Fisica, Univesit\'a di Roma  ``Sapienza'', p.le Aldo Moro 5, 00185, Roma, Italy.}} 
  \affiliation{Dipartimento di Fisica, Universit{\`a} di Trento, via Sommarive 14, 38050 Povo, Trento,
  Italy.}
  \affiliation{INFM CRS-SOFT, c/o Universit\`a di Roma ``Sapienza'', 00185, Roma, Italy.}

\author{T. S. Grigera} 
  \affiliation{Instituto de Investigaciones Fisicoqu{\'\i}micas
  Te{\'o}ricas y Aplicadas (INIFTA) and Departamento de F{\'\i}sica,
  Facultad de Ciencias Exactas, Universidad Nacional de La Plata,
  and CCT La Plata, Consejo Nacional de Investigaciones
  Cient{\'\i}ficas y T{\'e}cnicas, c.c. 16, suc. 4, 1900 La Plata, Argentina.}

\author{G. Parisi} 
  \affiliation{Dipartimento di Fisica, Universit{\`a} di Roma
  ``Sapienza'', p.le Aldo Moro 5, 00185, Roma, Italy. }
  \affiliation{Centre for Statistical Mechanics and Complexity (SMC), CNR-INFM.}

\author{P. Verrocchio} 
  \affiliation{Dipartimento di Fisica, Universit{\`a} di Trento, via Sommarive 14, 38050 Povo, Trento,
  Italy.}
  \affiliation{INFM CRS-SOFT, c/o Universit\`a di Roma ``Sapienza'', 00185, Roma, Italy.}
  \affiliation{Instituto de Biocomputaci\'on y F\'{\i}sica de Sistemas
  Complejos (BIFI), Spain.}

%% ABSTRACT %%%%%%%%%%%%%%%%%%%%%%%%%%%%%%%%%%%%%%%%%%%%%%%%%%%%%%%%%%%%%

\begin{abstract}
  We study dynamic heterogeneities in a model glass-former whose
  overlap with a reference configuration is constrained to a fixed
  value. The system phase-separates into regions of small
  and large overlap, so that dynamical correlations remain strong even
  for asymptotic times. We calculate an appropriate thermodynamic
  potential and find evidence of a Maxwell's construction consistent
  with a spinodal decomposition of two phases. Our results suggest
  that dynamic heterogeneities are the expression of an ephemeral
  phase-separating regime ruled by a finite surface tension.
\end{abstract}


\maketitle

%%
%% Acronyms (needs to go after \maketitle, at least for the twocolumns option)
%%

\acrodef{IS}{inherent structure}
\acrodef{RFOT}{random first-order theory}
\acrodef{CRR}{cooperatively rearranging regions}
\acrodef{MCT}{mode-coupling theory}
\acrodef{MC}{Monte Carlo}
\acrodef{KCM}{kinetically constrained models}

%% Article start %%%%%%%%%%%%%%%%%%%%%%%%%%%%%%%%%%%%%%%%%%%%%%%%%%%%%%%%%%%%%%

%%
%% Intro
%%

% xi dinamica
The conspicuous lack of a growing correlation length, contrasting with the
very steep increase of the relaxation time, has been a puzzle in the
physics of structural glasses
for quite a long time. Arguably, the first breakthrough has
been the discovery of dynamic heterogeneities \cite{review:Ediger00},
and the detection of a growing dynamical correlation
length, $\xd$ \cite{donati99,donati02}.  If we take two snapshots of
the system separated by a time lag comparable to the $\alpha$
relaxation time, $\tau_\alpha$, the particle displacements vary
enormously across the system, and the typical size $\xd$ of the
mobility-correlated regions increases on lowering the temperature.

% xi statica
More recently, by studying the thermodynamics of systems subject to
amorphous boundary conditions \cite{mosaic:bouchaud04,
  dynamics:montanari06}, an entirely different,
fully static, correlation length $\xs$ has been discovered
\cite{self:prl07,self:nphys08}.  $\xs$ also grows upon cooling, even
though its surge occurs at lower temperatures than $\xd$. The static
correlation  length has a natural interpretation as the size of the 
cooperatively rearranging regions 
\cite{glass:gibbs58}, and within the random first-order theory  
\cite{kirkpatrick89} it is determined by the balance between a surface tension cost 
and a configurational entropy gain of a rearrangement.

% mancanza di unificazione e nostro obiettivo
Clearly, it would be desirable to unify the dynamic and the
thermodynamic frameworks, so as to understand the interplay between
the two correlation lengths. Although the static-dynamic connection
is clear in mean-field systems \cite{self:castellani05} and some
progresses have been made in more realistic systems
\cite{Montanari06}, we are quite far from a unifying picture in real
glass-formers. Here we show that surface tension, which is a
crucial ingredient of the thermodynamic framework, also plays a key
role in the formation of dynamic heterogeneities. In so doing, we
establish a much-needed further link between dynamic and thermodynamic
relaxation in glass-forming liquids.

\begin{figure}
\includegraphics[width=8truecm]{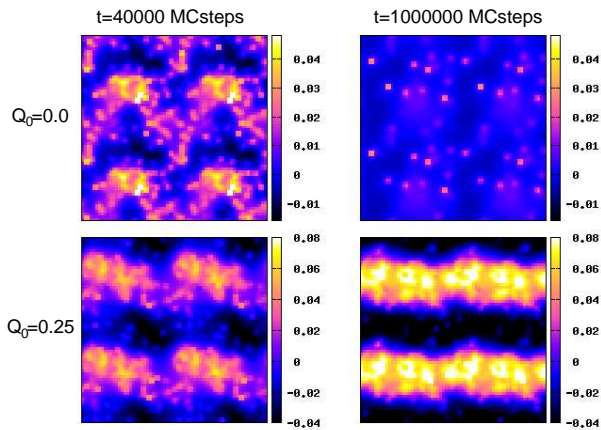}
\caption{Fluctuations of the overlap field, $\delta
  q(\mathbf{r},t)=q(\mathbf{r},t)-\langle q(t) \rangle $ for a 2-$d$
  slice of the system. Upper panels: unconstrained system. Lower
  panels: constrained system ($\hat Q=0.25$). Left panels:
  $t=\tau_\alpha$. Right panels: large times. $L=16$.}
\label{fig:eth}
\end{figure}

\begin{figure}
\includegraphics[angle=270,width=\columnwidth]{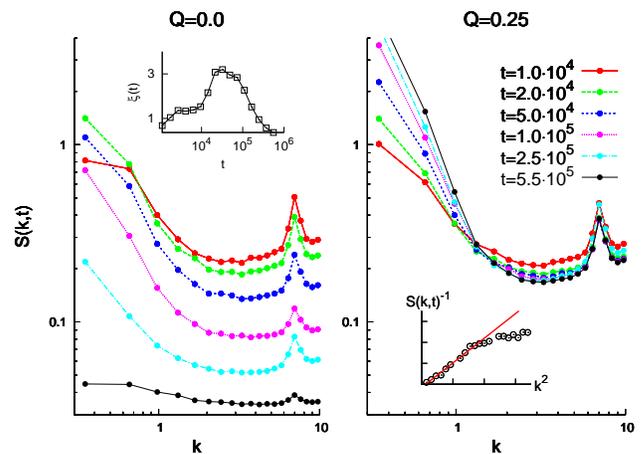}
\caption{$S(k,t)$ at different times for the unconstrained (left)
  and constrained (right, ($\hat Q = 0.25$) cases.
  Left inset: correlation length $\xi(t)$ as extracted from Eq.~\eqref{OZ}.
  Right inset: $S(k,t)^{-1}$ vs.\ $k^{2}$.  $T = \TMC$ and $L=16$.}
\label{fig:s4}
\end{figure}

% Le eterogeneita' nel caso standard
Let us start with the standard measurement of the dynamic
correlation length $\xd$.
% Il modello soft-spheres
Our glass-former is the well-known soft-sphere
model in 3-$d$ \footnote{We simulate the 3-$d$ soft-sphere binary mixture
  \cite{soft-spheres:bernu87} with parameters as in
  ref.~\onlinecite{self:nphys08}. Simulations were done with
  Metropolis Monte Carlo with particle swaps
  \cite{algorithm:Grigera01}. The mode-coupling temperature for this
  system is $\TMC=$0.226 \cite{soft-spheres:roux89}. Our largest
  system has $N=16384$ particles in a box of length $L=25.4$.  }.
% definizione Q(t) e q(x,t)
A useful tool to measure $\xd$ is the overlap, which quantifies how
much a configuration at time $t$ is similar to the reference
configuration at $t=0$. If we partition the system in small cubic
boxes and let $n_i$ be the number of particles in box
$i$, the local overlap is defined as
$q(\mathbf{r}_i,t)\equiv n_i(t) \, n_i(0)$, where $\mathbf{r}_i$
refers to the centre of cell $i$
\footnote{The side $\ell$ of the cells is such that the
  probability of finding more than one particle in a single box is
  negligible.}.
% Snapshots free
The spatial map of the local overlap tells us how much different regions
of the system have decorrelated (respect to the initial configuration)
over a time $t$.  In Fig.~\ref{fig:eth} (top) we show two snapshots of the overlap field. We
 see that at $t=\tau_\alpha$ there are large
heterogeneous regions, which eventually fade away for longer times.
% Def di G4 e S(k)
To quantify their size we must compute the overlap correlation function,
\begin{equation}
G(\mathbf{r},t) \equiv \big\langle q(0,t) q(\mathbf{r},t) \big\rangle -
\big\langle q(0,t) \rangle  \big\langle q(\mathbf{r},t) \big\rangle  ,
\label{G4}
\end{equation}
or its Fourier transform $S(k,t)$ (Fig.~\ref{fig:s4}, left).
In general, given a correlation function in
Fourier space, it is well-established practice \cite{caracciolo} 
to extract the correlation length $\xi$ from the
small-$k$ linear interpolation of $S^{-1}$ vs.\ $k^2$,
\begin{equation}
S(k,t)^{-1}= A + B  k^2  , 
\label{OZ}
\end{equation}
from which the correlation length is obtained as $\xi(t)^{2}= B/A$.  The
validity of Eq.~\ref{OZ} is shown in the inset of Fig.~2, right.
\footnote{The so-called {\it second-moment}
  correlation length is obtained by computing $A$ and $B$ using only
  the first two points, $\xi^2=[S^{-1}(k_1)/S^{-1}(0)-1]/k_1^2$,
  where $k_1=2\pi/L$ \cite{caracciolo}. However, we find that a linear
  fit to a few small-$k$ points gives equivalent results for
  $\xi^2$ yet lowering statistical errors. Of course, the
  free-field form \eqref{OZ} of $S(k,t)$ does not hold at generic
  values of $k$.}.  The time-dependent correlation lenght $\xi(t)$
represents the size of the dynamical heterogeneities at time $t$. This
lengthscale grows as the time approaches $\tau_\alpha$
and the heterogeneities become more extended (inset of
Fig.~2, left). The largest value of $\xi(t)$ (reached at the $\tau_\alpha$)
defines the so-called dynamical correlation length,
$\xd\equiv\xi(\tau_\alpha)$ \cite{donati99, donati02, LACEVIC03}.

% Xi e S(k) per tempi lunghi
What happens beyond $\tau_\alpha$? The memory of the initial
configuration is gradually lost, so that the correlation
function $S(k,t)$ decays sharply, Fig.~2 (left).  What happens to
$\xi(t)$ is less clear, because the vanishing of $S(k,t)$ makes it
hard to fit a reliable value of $\xi(t)$ through
Eq.~\ref{OZ}. Although this point is debated \cite{toninelli05}, our
results indicate that $\xi(t)$ decreases beyond $\tau_\alpha$ (left inset
of Fig.~2), in line with other studies \cite{LACEVIC03}.  However,
what is important, and definitely out of question, is that the {\it
  correlation} decreases for large times, irrespective of its spatial
range. Heterogeneities blur as $q(\mathbf{r},t)$ becomes
zero everywhere, and their size $\xi(t)$ becomes somewhat ill-defined.

% Il caso constrained
Let us now make a different experiment. We want to impose a constraint
on the dynamics, so that the system cannot
entirely lose memory of its initial configuration. This can be 
implemented by imposing a lower bound on the global overlap,
$Q(t) = 1/V \int d \mathbf{r} \; q(\mathbf{r},t)$.
In the unconstrained case $Q(t)$ goes asymptotically to zero, as the
memory of the initial configuration fades \footnote{Actually,
  normalization is such that $Q=\ell^3=0.062876$ for completely
  uncorrelated configurations, while $Q=1$ for two identical
  configurations.}.  On the other hand, if we run the dynamics with
the constraint $Q(t) \ge \hat Q$ things change \footnote{To enforce the
  constraint we modify the Metropolis algorithm: the probability to
  accept a move is $p = \min \{ 1,  \exp^{-\Delta E/T} \}$ for
  $ Q(t) \ge \hat Q$, and $p = 0$ for $Q(t) < \hat Q$. }. Initially
the system does not feel the constraint: the global overlap decreases
from its $t=0$ value, $Q=1$, and everything proceeds as described
above, including the growth of the heterogeneities.  However, at
later times $Q(t)$ hits its lower bound $\hat Q$ and it
cannot decrease further.  What happens to the dynamical
heterogeneities in this case?

% Le due ipotesi
There are two alternative hypotheses. First, the correlation $S(k,t)$
and its spatial range $\xi(t)$ decay to zero for large $t$ as in the
free case. Due to the constraint, however, such endgame cannot happen
in the same manner as in the free case, i.e.\ with $q(\mathbf{r},t)$
becoming zero everywhere. Heterogeneities must thus become very small,
forming a salt-and-pepper configuration of the field
$q(\mathbf{r},t)$, so that the total integral of the field stays equal to
$\hat Q$. Yet, if there is a nonzero surface tension between high and
low overlap regions, such a scenario is not what we expect: the
surface tension would force different domains to merge, driving the
system towards a phase-separated, highly correlated state
\cite{rev:bray94}. Hence, the second hypothesis is that the correlation {\it
  does not} decay and that the dynamic correlation length $\xi(t)$
grows beyond $\xd$, up to an asymptotic value of the order of the system's
size $L$. The stark difference between these two hypotheses
suggests that the constrained experiment may clarify the mechanisms of
formation of dynamical heterogeneities.

% Results: phase-separation
Inspection of the overlap field in the constrained case,
Fig.~\ref{fig:eth} (bottom), is quite telling: for large $t$ the system
phase separates into high and low overlap regions, forming stable
dynamical heterogeneities of the order of the system size. From a
quantitative point of view the situation is equally clear: in contrast
with the free case, the constrained correlation function does not go
to zero for large times, but saturates at a finite value,
Fig.~\ref{fig:s4} (right).  Hence, even in the late time regime
dynamic heterogeneities remain strongly correlated.

\begin{figure}
\includegraphics[angle=270,width=8truecm]{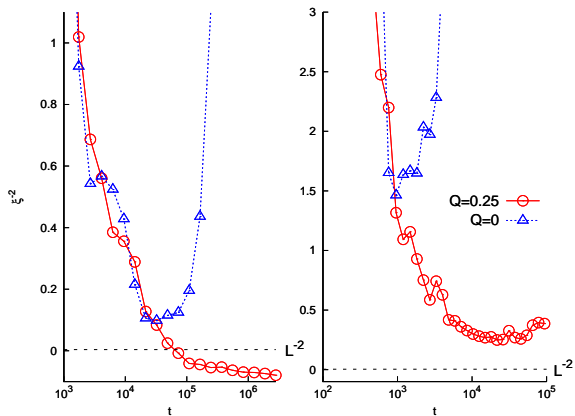}
\caption{Left: $A/B=\xi^{-2}(t)$ (see Eq.~\ref{OZ}) at $T= \TMC$
  in the constrained (circles) and unconstrained (triangle)
  cases. Right: the same at $T= 1.55 \TMC$. $L=16$.}
\label{fig:xi}
\end{figure}

% Parte cazzuta su \xi^{-2}
The study of the correlation length in the constrained case confirms
this scenario \footnote{Due to the constraint, the space integral of
  the correlation function is zero, hence the single point
  $S^{-1}(0,t)= [\int d\mathbf r\; G(\mathbf r,t)]^{-1}$ must be
  excluded from the analysis. This also
  implies that the dynamical susceptibility, $\chi(t)= S(0,t)=
  V[\langle Q^2(t)\rangle -\langle Q(t)\rangle^2]$, a standard marker
  of heterogeneous dynamics, is trivially zero and therefore
  useless with the constraint.}.  In an infinite system undergoing
phase separation, or at a critical point, the intercept $A$ in
Eq.~\ref{OZ} vanishes while the slope $B$ remains finite, so that
$\xi$ grows indefinitely \cite{caracciolo}. On the other hand, in a
      {\it finite} system phase separation means that $\xi$ becomes
      comparable with system size $L$. As a consequence, the
      finite-size (periodic) real space correlation function
      $G(\mathbf r,t)$ ceases to be a simple exponential for large $r$
      (small $k$). This implies that the intercept $A$ in Eq.~\ref{OZ}
      can go below zero and take small negative values
      ($O(1/L^2)$). In a finite-size system it is therefore convenient
      to compare $A/B=\xi^{-2}$ vs.\ $L^{-2}$ to check whether or not
      phase separation occurs. From Fig.~\ref{fig:xi} (left) we see that
      in the unconstrained case $\xi^{-2}$ keeps well clear of
      $L^{-2}$, while in the constrained case $\xi^{-2}(t)$
      unmistakably goes below $L^{-2}$. This is exactly what we expect in
      a system with nonzero surface tension undergoing phase
      separation. We studied two other sizes, $L=8$ and $L=25$, and in
      both cases $\xi^{-2}(t)$ drops below $L^{-2}$, indicating phase
      separation.
% xi(t) vs t ad alta temperatura
At higher temperatures, however, though the correlation is
enhanced by the constraint, the latter is ineffective to make
$\xi^{-2}(t)$ drop below $L^{-2}$ (Fig.~\ref{fig:xi}, right). These
results are consistent with the idea that the surface tension decays
at high temperature, thus preventing phase separation
\cite{self:Cammarota09a}.

\begin{figure}
\includegraphics[angle=270,width=7truecm]{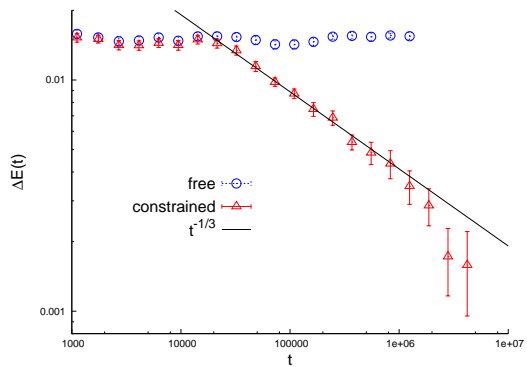}
\caption{Energy difference $\Delta E(t)=E(t)-E_{0}$ vs
  $t$ at $T = \TMC$ with constrained dynamics, $\hat Q=0.25$.
  $E_{0}$ is a parameter of the fit $E(t)=E_{0}+\gamma t^{-1/3}$.
  The line is $1/t^{1/3}$, corresponding to the  surface tension exponent
  $\theta=2$.  }
\label{fig:ene}
\end{figure}

% Coarsening and the Energy
In systems with conserved order parameter undergoing phase separation
the domains size $\xi(t)$ grows as $t^{1/3}$ and the dynamics proceeds
by reducing the total amount of interfaces, and therefore of energy, in the
system \cite{rev:bray94}.  The interface energy per domain scales like
$\xi^\theta$, where $\theta$ is the surface tension exponent. The
total number of domains is $L^d/\xi^d$, so that the total interface
energy density is $\Delta E(t) \sim 1/\xi(t)^{d-\theta} \sim
1/t^{(d-\theta)/3}$. In the standard case $\theta=d-1$, so that
$\Delta E(t) \sim 1/t^{1/3}$ \cite{rev:bray94}. Fig.\ref{fig:ene}
shows that something remarkably similar happens in  our case. After
the constraint kicks in, $\Delta E(t)$ decays compatibly with an
exponent $1/3$.  Hence, even though fitting coarsening exponents
is notoriously difficult, and one must be careful in drawing any
conclusion, our data seems to be compatible with the `naive' exponent
$\theta=2$ \cite{self:Cammarota09a, self:Cammarota09b}.

% Potential: motivation 
In general, phase-separation is the landmark of first order phase transitions and metastability. At the
mean-field level one can normally define a thermodynamic potential as a function of the
order parameter that, below some spinodal point, exhibits a stable and a
metastable minimum, corresponding to the two phases. In
finite dimension Maxwell's construction makes the potential convex, so
that the derivative of the potential is constant (zero second
derivative) in a finite interval (Fig.~5, inset). Maxwell's construction
implies that when the order parameter is conserved and constrained 
to take a value in the non-convex interval, phase separation
occurs. We have clearly observed phase-separation.  Can we define a
thermodynamic potential displaying Maxwell's construction?

\begin{figure}
\includegraphics[angle=270,width=8truecm]{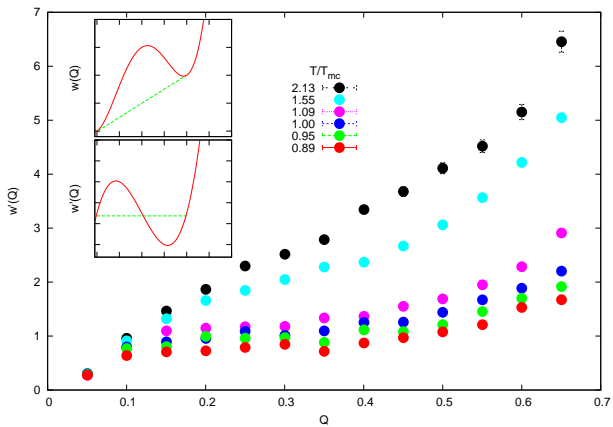}
\caption{The derivative $W^\prime(\hat Q)$ of the thermodynamic potential at different
temperatures, ranging from $2.13  \TMC$ to $0.89  \TMC$. Inset: a cartoon of 
Maxwell's construction for the potential and its derivative.}
\label{fig:wprimo}
\end{figure}

% Potential: definition
Let us proceed minimalistically. Our phase-separating order parameter
is the overlap $Q$, so it is a potential $W(Q)$ we are after.
Besides, the potential must determine the observed probability
distribution of $Q$ through the relation, $P(Q) = \exp \left[-N
  W(Q)\right] \, \theta(Q-\hat Q)$, where the $\theta$-function
enforces the constraint \cite{Parisi09}. If we compute the average 
linear fluctuation of $Q$ and expand the exponential, we obtain,
\begin{equation}
W'(\hat Q) \sim N^{-1}\ \langle Q-\hat Q  \rangle^{-1}  . 
\end{equation}
This quantity is easy to compute: we let the system evolve until the
constraint is hit, and then we measure the (very small) average
fluctuation of the overlap $Q$ over $\hat Q$ (see
\cite{algorithm:Fernandez09} for a different definition of the potential).
We report $W^\prime(\hat Q)$ in Fig.(\ref{fig:wprimo}). The second
derivative of the potential is clearly nonzero at high $T$, whereas
around the Mode Coupling temperature a finite region with $W^{\prime
  \prime}(\hat Q) \sim 0$ develops.  This is evidence of Maxwell's
construction and it supports the link between phase separation and
metastability in our system.

% Commenti sul V2(q)
$W(Q)$ is a finite-dimensional
variant of the two-replica potential originally introduced in
mean-field spin-glasses \cite{franz-parisi-95}, and later generalized
to structural glasses \cite{franz-parisi-98}.  This potential is the
free energy cost to keep a configuration (the running one in the
present work) at fixed overlap $Q$ with a generic equilibrium
configuration (the initial reference one). Below a dynamic transition
(roughly, the Mode Coupling temperature), the mean-field potential
develops a metastable minimum at a finite value of $Q$. In this
framework relaxation at low temperatures can be interpreted as a
barrier crossing process, bringing the system from the metastable
minimum (short times, finite $Q$) to the stable minimum (long times, zero
$Q$) \cite{nucleation:franz05}. The constraint is just a stratagem
to keep the overlap within the nonconvex region of the
potential, as to interrupt relaxation and therefore force phase
separation.

%% Conclusions
We have studied dynamic heterogeneities in a glass-forming liquid with
constrained global overlap. At low temperature both the dynamic correlation
function  and the thermodynamic potential indicate that
the system phase separates into regions of high and low overlap.  On
the contrary, at high temperature no phase separation occurs,
supporting the view of a surface tension that decreases at high
$T$. The co-existence of regions belonging to different amorphous
`states' (here the high/low overlap patches) is reminiscent of the
random first-order theory of thermodynamic relaxation \cite{kirkpatrick89}. In
the dynamical case the evolution of these regions is driven by a
classic coarsening mechanism, which is stable with the constraint,
but ephemeral in absence of the constraint. In the
thermodynamic case, on the other hand, the evolution of these regions
is presumably driven by an entropic mechanism
\cite{kirkpatrick89}. Our results show that surface tension and
metastability stand as key links between the two frameworks.

% acknowledgements
We thank G.~Biroli, J.-P.~Bouchaud, L.~Cugliandolo, S.~Franz, W.~Kob
and F.~Zamponi for several important remarks, and ECT* and CINECA for
computer time. The work of TSG was supported in part by grants from
ANPCyT, CONICET, and UNLP (Argentina).

%% Bibliography %%%%%%%%%%%%%%%%%%%%%%%%%%%%%%%%%%%%%%%%%%%%%%%%%%%%%%%%%%%%%%%

\bibliography{references}
\bibliographystyle{apsrev}

\end{document}